\documentclass[iop,apjl]{emulateapj}
\newcommand{\V}[1]{\mathbf{#1}} 
\newcommand\Alfven{Alfv\'en } 
\usepackage{amssymb}
\usepackage{amsmath}
\usepackage{hyperref}
\usepackage{natbib}
\usepackage{color}
\usepackage{graphicx}
\usepackage{amsmath}
\usepackage{bm}

\begin{document}
\title{A Modified Version of Taylor's Hypothesis for Solar Probe Plus Observations}

\author{Kristopher ~G. Klein, Jean C. Perez, Daniel Verscharen, Alfred
  Mallet, \& Benjamin ~D.~G. Chandran$^1$}

\affiliation{$^1$Space Science Center, University of New Hampshire, 
Durham, NH 03824, USA\\}

\begin{abstract}
  The Solar Probe Plus (SPP) spacecraft will explore the near-Sun
  environment, reaching heliocentric distances less than
  $10 R_{\odot}$.  Near Earth, spacecraft measurements of fluctuating
  velocities and magnetic fields taken in the time domain are
  translated into information about the spatial structure of the solar
  wind via Taylor's ``frozen turbulence'' hypothesis.  Near the
  perihelion of SPP, however, the solar-wind speed is comparable to
  the Alfv\'en speed, and Taylor's hypothesis in its usual form does
  not apply.  In this paper, we show that, under certain assumptions,
  a modified version of Taylor's hypothesis can be recovered in the
  near-Sun region. We consider only the transverse, non-compressive
  component of the fluctuations at length scales exceeding the proton
  gyroradius, and we describe these fluctuations using an approximate
  theoretical framework developed by Heinemann and Olbert. We show
  that fluctuations propagating away from the Sun in the plasma frame
  obey a relation analogous to Taylor's hypothesis when
  $V_{\rm sc,\perp} \gg z^-$ and $z^+ \gg z^-$, where
  $V_{\rm sc,\perp}$ is the component of the spacecraft velocity
  perpendicular to the mean magnetic field and $\bm{z}^+$ ($\bm{z}^-$)
  is the Elsasser variable corresponding to transverse,
  non-compressive fluctuations propagating away from (towards) the Sun
  in the plasma frame.  Observations and simulations suggest that, in
  the near-Sun solar wind, the above inequalities are satisfied and
  $\bm{z}^+$ fluctuations account for most of the fluctuation energy.
  The modified form of Taylor's hypothesis that we derive may thus
  make it possible to characterize the spatial structure of the
  energetically dominant component of the turbulence encountered
  by~SPP.
\end{abstract}


\maketitle

\section{Introduction}
\label{sec:intro}

Understanding the processes that heat the Sun's corona and accelerate
the solar wind is a long standing goal in the space-physics community.
A number of mechanisms have been proposed to account for these
phenomena, including low-frequency Alfv\'en-wave turbulence
\citep{Coleman:1968,velli89}, cyclotron heating
\citep{Ionson:1978,hollweg02}, stochastic heating
\citep{mcchesney87,chaston04,Chandran:2010b}, velocity filtration
\citep{Scudder:1992a}, reconnection \citep{Parker:1972}, and
nanoflares \citep{Parker:1988}. Remote observations have been employed
to constrain the likelihood of these mechanisms operating near the Sun
\citep{hollweg82b,Harmon:2005,DePontieu:2007,Tomczyk:2007}. However,
\emph{in situ} measurements are necessary to distinguish between these
competing ideas.

The upcoming Solar Probe Plus (SPP) mission will make such
measurements at heliocentric distances~$r$ as small as~$9.87
R_{\odot}$, where $R_{\odot}$ is the Solar radius.  In order
to extract scientific results from SPP measurements, we must consider
the interpretation of measurements made in the spacecraft reference
frame.  Spacecraft measurements are made in the time domain, but are
often translated into information about the spatial structure of waves
and turbulence using Taylor's hypothesis
\citep{Taylor:1938,Fredricks:1976}, 
which treats fluctuations as
static in the reference frame that moves with the mean
velocity~$\bm{U}$ of the solar wind (which we take to be measured in
an inertial reference frame centered on the Sun).  For example, if
$\bm{B}_{\rm sc}(t)$ denotes the magnetic field measured by the
spacecraft and $\bm{B}_{\rm SW}(\bm{x})$ denotes the (assumed-to-be)
static magnetic field as a function of position~$\bm{x}$ in the
solar-wind frame, then
\begin{equation}
\bm{B}_{\rm sc}(t) = \bm{B}_{\rm SW}(\bm{x}_0 - \bm{U}t),
\label{eq:Taylor1} 
\end{equation} 
where $\bm{x}_0$ is the spacecraft location 
at~$t=0$. Here, we have neglected the velocity of the spacecraft,
which, near Earth, is $\ll U$. Taylor's hypothesis is a good
approximation when $U$ is much larger than the fluctuating plasma
velocity
and wave phase speeds, since waves and
turbulent structures are then advected past the spacecraft on a
timescale that is much shorter than the time required for the waves or
structures to evolve appreciably in the solar-wind frame.

When Taylor's hypothesis holds and~$\bm{U}$ is constant, the (angular)
frequency power spectrum $P_\omega(\omega)$ of a quantity such as
$\bm{B}_{\rm sc}(t)$ is related to the wavenumber spectrum
$P_{\rm 3D}(\bm{k})$ of $\bm{B}_{\rm SW}(\bm{x})$ through the
equation~\citep{Jokipii:1973,horbury08,bourouaine13a}
\begin{equation}
P_\omega(\omega) = \int P_{\rm 3D} (\bm{k}) \delta (\bm{k} \cdot \bm{U} - \omega) d^3 k,
\label{eq:PfP3D} 
\end{equation} 
where the $k$ integration is over all of $k$-space.  Thus, a
wavenumber~$\bm{k}$ in the solar-wind frame corresponds to an 
angular frequency~$\omega = \bm{k} \cdot \bm{U}$ in the spacecraft frame.

A critical issue for SPP is that $U$ is comparable to the Alfv\'en
speed~$v_{\rm A}$ near SPP's perihelion, and thus Taylor's hypothesis
does not in general apply. However, in this paper, we show that a
modified version of Taylor's hypothesis can be recovered near SPP's
perihelion under a set of conditions that are expected to hold for the
energetically dominant component of the turbulent fluctuations in the
near-Sun solar wind.

\section{A Modified Taylor's Hypothesis for SPP}
\label{sec:HO}

In situ measurements indicate that the (total) fractional density
fluctuations~$\delta \rho/\rho_0$ in the solar wind at $r>60 R_{\sun}$
are much smaller than~$|\delta \bm{B}|/B_0$, where $\delta \bm{B}$ and
$\bm{B}_0$ are, respectively, the fluctuating and background magnetic
fields, and $\delta \rho$ and $\rho_0$ are the fluctuating and
background mass densities, respectively~\citep{tumarsch95}. Observations of radio
signals transmitted by the {\em Helios} spacecraft indicate that the
inequality $\delta \rho/\rho_0 \ll |\delta \bm{B}|/B_0$ also holds
near SPP's perihelion \citep[see, e.g., the appendix
of][]{Hollweg:2010}.  The condition
$\delta \rho/\rho_0 \ll |\delta \bm{B}|/B_0$ implies that the dominant
fluctuations are non-compressive, consistent with the fact that
compressive waves are damped much more rapidly than non-compressive
waves in the collisionless solar wind~\citep{Barnes:1966} and with
Hollweg's~(\citeyear{Hollweg:1978}) finding that fast magnetosonic
waves launched outward through the chromosphere are reflected almost
completely at the transition region.

In this paper, we restrict our analysis to the dominant,
non-compressive component of the fluctuations and derive a version of
Taylor's hypothesis that will apply only to such non-compressive
fluctuations. We assume that
\begin{equation}
\delta \rho \ll \rho_0.
\label{eq:drho} 
\end{equation} 
We also restrict our analysis to fluctuations at length
scales greater than the proton gyroradius.  At such
scales, the only non-compressive mode is the Alfv\'en wave, for which
$\delta \bm{B}$ and the fluctuating velocity~$\delta \bm{v}$ are
perpendicular to~$\bm{B}_0$.  We thus take the fluctuations to satisfy
the conditions
\begin{equation}
\nabla \cdot \delta \bm{v} = 0 \qquad
\delta \bm{v}\cdot\bm{B}_0 = 0 \qquad \delta \bm{B} \cdot \bm{B}_0 = 0.
\label{eq:TNC} 
\end{equation} 

We assume that $\bm{B}_0$ and the background solar-wind
velocity~$\bm{U}$ are, at least to a good approximation, in the radial
direction~$\bm{\hat{r}}$. (It is trivial to generalize our results to
the case in which $\bm{B}_0$ points in the $-\bm{\hat{r}}$ direction.)
We also take $\rho_0$ to vary with position perpendicular
to~$\bm{\hat{r}}$ much more slowly than does~$\delta \bm{B}$.  Given
these assumptions and Equations~(\ref{eq:drho}) and~(\ref{eq:TNC}),
the fluctuations are described by the
Heinemann-Olbert~(\citeyear{Heinemann:1980}) equations~\citep[for
additional discussion see
][]{Chandran:2009d},
\[
\frac{\partial \bm{g}}{\partial t} 
+ (\bm{U} + \bm{v}_{\rm A}) \cdot \nabla \bm{g} 
 - \left(\frac{{U} + {v}_{\rm A}}{2v_{\rm A}}\right) \frac{d v_{\rm A}}{d r}\,\bm{f}
\hspace{1.5cm} 
\]
\begin{equation} 
\hspace{1cm}  =
 - \bm{z}^- \cdot \nabla \bm{g}
- \left(\frac{1 + \eta^{1/2}}{\eta^{1/4}}\right) \frac{\nabla p_{\rm tot}}{\rho_0}
\label{eqn:HOg}
\end{equation}
and 
\[
\frac{\partial \bm{f}}{\partial t} 
+ (\bm{U} - \bm{v}_{\rm A}) \cdot \nabla \bm{f} 
-\left(\frac{{U} - {v}_{\rm A}}{2v_{\rm A}}\right) \frac{d v_{\rm A}}{d r}\,\bm{g}
\hspace{1.5cm} 
\]
\begin{equation} 
\hspace{1cm} 
= - \bm{z}^+ \cdot \nabla \bm{f}
- \left(\frac{1 - \eta^{1/2}}{\eta^{1/4}}\right) \frac{\nabla p_{\rm tot}}{\rho_0},
\label{eqn:HOf}
\end{equation}
where
\begin{equation}
\bm{g} =\left(\frac{1 + \eta^{1/2}}{\eta^{1/4}}\right) \bm{z}^+
\qquad
\bm{f} =\left(\frac{1 - \eta^{1/2}}{\eta^{1/4}}\right) \bm{z}^-
\label{eq:fg} 
\end{equation} 
are the Heinemann-Olbert variables, $\bm{v}_{\rm A} = \bm{B}_0/\sqrt{
  4 \pi \rho_0}$ is the Alfv\'en velocity, $p_{\rm tot}$ is the
total pressure,
\begin{equation}
\eta \equiv \frac{\rho_0}{\rho_a} = \frac{v_{\rm A}^2}{U^2},
\label{eqn:eta}
\end{equation}
$\rho_a$ is the value of~$\rho_0$ at the \Alfven critical point~$r=r_{\rm a}$
(at which~$U=v_{\rm A}$), 
\begin{equation} 
\bm{z}^\pm = \delta \bm{v} \mp \delta \bm{b}
\label{eqn:zz}
\end{equation} 
are the Elsasser variables, and $\delta \bm{b} = \delta
\bm{B}/\sqrt{4 \pi \rho_0}$.  Physically, $\bm{g}$ and $\bm{z}^+$
($\bm{f}$ and $\bm{z}^-$)
represent noncompressive, transverse fluctuations that propagate away
from (toward) the Sun in the plasma frame.

The pressure terms in Equations~(\ref{eqn:HOg}) and (\ref{eqn:HOf})
enforce the incompressibility condition by canceling out the
compressive components of the nonlinear terms $\bm{z}^+ \cdot \nabla
\bm{f}$ and $\bm{z}^-\cdot \nabla \bm{g}$. To simplify the notation,
we define
\begin{equation}
\left( \bm{z}^- \cdot \nabla \bm{g}\right)_{\rm nc} = 
\bm{z}^- \cdot \nabla \bm{g} + 
\left(\frac{1 + \eta^{1/2}}{\eta^{1/4}}\right) \frac{\nabla p_{\rm tot}}{\rho_0},
\label{eq:nlnc} 
\end{equation} 
where the subscript ``nc'' stands for ``non-compressive component.''

Measurements from the {\em Helios} spacecraft 
show that on average
the ratio $z^+/z^-$ exceeds~1 at $0.3 \mbox{ AU} < r <
1 \mbox{ AU}$ 
 and that this ratio
increases as $r$ decreases~\citep{Bavassano:2000}. Theoretical
models~\citep{Cranmer:2005,Verdini:2007,Chandran:2009d}, shell-model
simulations~\citep{Verdini:2012}, and direct numerical
simulations~\citep{Perez:2013} also suggest that
$z^+ \gg z^-$ at $r\sim 10 - 30 R_{\sun}$. We thus assume that
\begin{equation}
z^- \ll z^+.
\label{eq:zmzp} 
\end{equation} 
Equation~(\ref{eq:zmzp}) implies that $f \ll g$. Moreover, the ratio
$f/g$ is much smaller than $z^-/z^+$ near~$r=r_{\rm a}$, because of
the prefactor~$(1-\eta^{1/2})/\eta^{1/4}$ in the definition
of~$\bm{f}$ in Equation~(\ref{eq:fg}).  The quantity~$(1/v_{\rm A})dv_{\rm
  A}/dr$ is $\sim r^{-1}$.  It thus follows from Equation~(\ref{eq:zmzp})
that $|(\bm{U}+ \bm{v}_{\rm A} ) \cdot \nabla \bm{g} |\gg | [({U} +
  {v}_{\rm A})/(2v_{\rm A})] (d v_{\rm A}/d r)\,\bm{f}|$, because
$|\partial \V{g}/\partial r|$ is not much smaller than~$|g/r|$.
Given Equation~(\ref{eq:zmzp}), we may thus approximate 
Equation~(\ref{eqn:HOg}) as
\begin{equation}
\frac{\partial \bm{g}}{\partial t} 
+ (\bm{U} + \bm{v}_{\rm A}) \cdot \nabla \bm{g}  =
 - \left(\bm{z}^- \cdot \nabla \bm{g}\right)_{\rm nc}.
\label{eqn:HOg2}
\end{equation}

We now change from a reference frame centered on the Sun with
position~$\bm{r}$ and time~$t$ to the spacecraft reference frame with
position $\bm{r}' = \bm{r} -\int\bm{V}_{\rm sc}dt$ and time $t' = t$.
The temporal and spatial derivatives in the two frames are related by
the equations
$\partial/\partial t = \partial/\partial t' + \partial
\bm{r}'/\partial t \cdot \nabla' $
and $\nabla = \nabla'$, where
$\partial \bm{r}'/\partial t = -\bm{V}_{\rm sc}$. We can thus rewrite Equation~(\ref{eqn:HOg2}) as
\begin{equation}
\frac{\partial \bm{g}}{\partial t'} 
+ (\bm{U} + \bm{v}_{\rm A} - \bm{V}_{\rm sc}) \cdot \nabla' \bm{g} =
 - \left(\bm{z}^- \cdot \nabla' \bm{g}\right)_{\rm nc},
\label{eqn:HOgsc}
\end{equation}
where $\left(\bm{z}^- \cdot \nabla' \bm{g}\right)_{\rm nc}$ is given by 
the right-hand side of Equation~(\ref{eq:nlnc}), with~$\nabla$
replaced by $\nabla'$.

Our goal now is to determine when the nonlinear term on the right-hand
side of Equation~(\ref{eqn:HOgsc}) can be neglected.  We note that the
$\bm{z}^- \cdot \nabla' \bm{g}$ term ``picks up'' the spatial
derivatives of~$\bm{g}$ in the directions perpendicular to~$\bm{B}_0$,
because of Equation~(\ref{eq:TNC}).  On the other hand, the $(\bm{U} +
\bm{v}_{\rm A})\cdot \nabla' \bm{g}$ term picks up the spatial
derivatives of~$\bm{g}$ along the direction of~$\bm{B}_0$. Thus, even
though $U+v_{\rm A} \gg z^-$, it is not necessarily the case that
$|(\bm{U} + \bm{v}_{\rm A})\cdot \nabla' \bm{g}| \gg |\bm{z}^-
\cdot\nabla' \bm{g}|$, because the perpendicular gradient of~$\bm{g}$
could greatly exceed the parallel gradient. On the other hand, the
spacecraft velocity~$\bm{V}_{\rm sc}$ has a nonzero component
$\bm{V}_{\rm sc, \perp} \equiv \bm{V}_{\rm sc} - \hat{b} (\hat{b}
\cdot \bm{V}_{\rm sc}) $ perpendicular to~$\bm{B}_0$.  The term
$\bm{V}_{\rm sc} \cdot \nabla'\bm{g}$ thus greatly exceeds $\bm{z}^-
\cdot \nabla' \bm{g}$ in magnitude when
\begin{equation}
V_{\rm sc,\perp} \gg z^-.
\label{eq:vsczm} 
\end{equation} 
Near perihelion, $V_{\rm sc,\perp} \simeq 200 \mbox{ km/s}$
and Equation~(\ref{eq:vsczm}) likely holds
\citep{Cranmer:2005,Verdini:2007,Chandran:2009d}.
When Equation~(\ref{eq:vsczm}) is satisfied, 
Equation~(\ref{eqn:HOgsc}) becomes (to leading order in
the small quantity $z^-/V_{\rm sc,\perp}$)
\begin{equation}
\frac{\partial \bm{g}}{\partial t'} +
\bm{U}_{\rm total}\cdot \nabla' \bm{g}=0,
\label{eqn:HOg3}
\end{equation}
where
\begin{equation} 
\bm{U}_{\rm total} \equiv \bm{U} + \bm{v}_{\rm A} - \bm{V}_{\rm sc}.
\label{eq:defUtot} 
\end{equation} 
The left-hand side of Equation~(\ref{eqn:HOg3}) is the convective time
derivative of $\bm{g}$ at a point that moves with
velocity~$\bm{U}_{\rm total}$.  The vanishing of this time derivative
to leading order in $z^-/V_{\rm sc,\perp}$ and $z^-/z^+$ expresses
mathematically the statement that $\bm{g}$ fluctuations
are approximately frozen in a frame that moves with
velocity~$\bm{U}_{\rm total}$.

We now restrict our consideration to time intervals $\tau$ that are
sufficiently short that (1) $\bm{V}_{\rm sc}$ is approximately constant
and (2) $\bm{v}_{\rm A}$ and $\bm{U}$ are approximately constant at the
spacecraft location and throughout the radial interval through which
the $\bm{g}$ fluctuations propagate during time~$\tau$.  With these
conditions, $\bm{U}_{\rm total}$ can be treated as constant, and the
solution for $\bm{g}$ in Equation~(\ref{eqn:HOg3}) is
\begin{equation}
\bm{g}(\bm{r}',t') = \bm{g}\left(\bm{r}' - \bm{U}_{\rm total}t',0\right).
\label{eqn:thHO}
\end{equation} 
Just as Equation~(\ref{eq:Taylor1})  leads to Equation~(\ref{eq:PfP3D}),
Equation~(\ref{eqn:thHO})  implies that
\begin{equation}
P_\omega^{(g)}(\omega) =  \int P_{\rm 3D}^{(g)} (\bm{k}) \delta (\bm{k} \cdot \bm{U}_{\rm total} - \omega) d^3 k,
\label{eq:PfP3D2} 
\end{equation} 
where $P_\omega^{(g)}$ is the (angular) frequency spectrum of 
$\bm{g}$ in the spacecraft frame, and $P_{\rm 3D}^{(g)}$ is
the wavenumber spectrum of~$\bm{g}(\bm{r}',0)$.  In order for the
inward-propagating fluctuations (i.e., $\bm{f}$) to obey an analogous
relation, $|\bm{V}_{\rm sc, \perp}|$ would need to be $\gg
|\bm{z}^+|$, which is not expected for the near-Sun environment.

\begin{figure}[t]
\centerline{
\includegraphics[width=5.5cm]{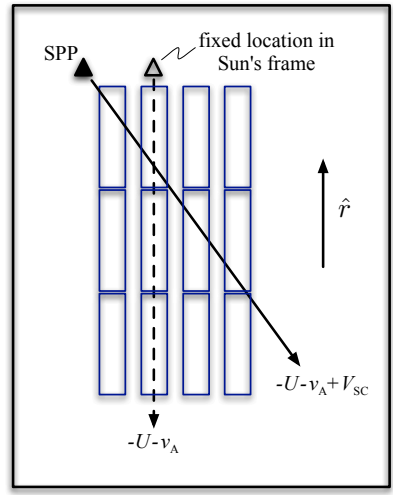}
}
\caption{Motion of SPP (filled triangle) in the ``$z^+$ rest frame,''
  which moves away from the Sun at velocity~$\bm{U} + \bm{v}_{\rm
    A}$. The open triangle represents a point that is at rest in an
  inertial frame centered on the Sun. The rectangles represent
  turbulent structures associated with non-compressive $z^+$ (or $g$)
  fluctuations.
\label{fig:spp} }
\end{figure}

To understand why a version of Taylor's hypothesis applies to
outward-propagating $\bm{g}$ fluctuations even when $U\sim v_{\rm A}$,
it is helpful to view the spacecraft in the ``$z^+$ reference frame,''
which moves away from the Sun at velocity $\bm{U} + \bm{v}_{\rm A}$,
which, as discussed above, we treat as as effectively constant. In the
$z^+$ reference frame, $\bm{g}$ changes only because of the nonlinear
shearing represented by the $-(\bm{z}^-\cdot\nabla \bm{g})_{\rm nc}$
term on the right-hand side of Equation~(\ref{eqn:HOg2}). A fixed
location in an inertial reference frame centered on the Sun moves at
velocity~$- \bm{U} - \bm{v}_{\rm A}$ in the $z^+$ frame, as
represented by the open triangle in Figure~\ref{fig:spp}. At such a
location, our modified version of Taylor's hypothesis (in this case
with $V_{\rm sc} = 0$) need not apply, because the timescale on which
a $\bm{g}$ structure evolves due to nonlinear shearing could be
comparable to the time needed to traverse that structure in the radial
direction at speed~$U+ v_{\rm A}$.  This is directly related to the
point made above that even though $U+v_{\rm A} \gg z^-$, it is not
necessarily the case that
$|(\bm{U} + \bm{v}_{\rm A})\cdot \nabla' \bm{g}| \gg |\bm{z}^-
\cdot\nabla' \bm{g}|$,
because $\bm{U} + \bm{v}_{\rm A}$ is along~$\bm{B}_0$, whereas
$\bm{z}^{-}$ is perpendicular to~$\bm{B}_0$.  On the other hand, in
the $z^+$ frame, the SPP spacecraft will move at velocity
$-\bm{U} -\bm{v}_{\rm A} + \bm{V}_{\rm sc}$, as illustrated by the
filled triangle in Figure~\ref{fig:spp}.  If Equation~(\ref{eq:vsczm})
is satisfied, then SPP will traverse a $\bm{g}$ structure in a time
that is much shorter than the time required for that structure to
change in the $z^+$ frame, and the $\bm{g}$ structures can be
approximated as frozen.

Equations~(\ref{eqn:thHO}) and (\ref{eq:PfP3D2}) are analogous to the
usual forms of Taylor's hypothesis given in
Equations~(\ref{eq:Taylor1}) and (\ref{eq:PfP3D}), but with some
differences.  At $r=1 \mbox{ AU}$, structures are approximately frozen
in the plasma frame and are swept by the spacecraft at a
velocity~$\simeq \bm{U}$.  Closer to the Sun, when
Equations~(\ref{eq:zmzp}) and (\ref{eq:vsczm}) are satisfied, $\bm{g}$
structures are approximately frozen and are swept by the spacecraft at
velocity~$\bm{U} + \bm{v}_{\rm A} - \bm{V}_{\rm sc}$, but $\bm{f}$
structures need not be frozen.  At $r=1 \mbox{ AU}$, the frequencies
in the spacecraft frame are much larger than the frequencies of
fluctuations measured in the plasma frame. In contrast, this frequency
disparity need not arise near the Alfv\'en critical point. If the
frequency associated with some fluctuation in the spacecraft frame
arises primarily from the
$(\bm{U} + \bm{v}_{\rm A})\cdot \nabla'\bm{g}$ term in
Equation~(\ref{eqn:HOgsc}), then this frequency is only a factor of
$(U+v_{\rm A})/v_{\rm A} \sim 2$ larger than the frequency that would
arise in the plasma frame.

The direct output of Equations~(\ref{eqn:thHO}) and (\ref{eq:PfP3D2})
is the spatial structure of the~$\bm{g}$ field. However, because we
restrict our analysis to time intervals~$\tau$ that are sufficiently
small that $U$ and $v_{\rm A}$ remain fairly constant at the
spacecraft location and throughout a radial interval extending from
SPP a distance ~$\sim (U + v_{\rm A})\tau$ toward the Sun, the factor
of proportionality between~$\bm{z}^+$and~$\bm{g}$,
\begin{equation}
h \equiv \frac{\eta ^{1/4}}{1 + \eta^{1/2}},
\label{eqn:H}
\end{equation}
can be treated as approximately constant. As a consequence, the
frequency and wavenumber spectra of~$\bm{g}$ in
Equation~(\ref{eq:PfP3D2}) can be converted into frequency and
wavenumber spectra of~$\bm{z}^+$ via Equation~(\ref{eq:fg}).
Near the Alfv\'en critical point $r_{\rm a}$, the radial variations in
$h$ are particularly small, allowing for a translation between
$\bm{g}$ spectra and $\bm{z}^+$ spectra with very little error.
This can be seen by writing $\eta = 1 + x$, where
$ |x| \ll 1$ near $r=r_{\rm a}$.  
From Equation~(\ref{eqn:eta}), $x= (v_{\rm A}^2 - U^2)/U^2$.
Taylor-expanding Equation~(\ref{eqn:H}) about $\eta=1$, we obtain
\begin{equation}
h = \frac{1}{2} \left[ 1 - \frac{(U^2 - v_{\rm A}^2)^2}{32U^4} + ...\right].
\label{eqn:Hx}
\end{equation}
Because $h$ varies quadratically with the quantity $(U^2 - v_{\rm
  A}^2)$, and because of the factor of $1/32$, $h$ varies very slowly
with radius near~$r_{\rm a}$.

To illustrate how the technique we describe could be applied, we
suppose that near its perihelion of~$9.87 R_{\sun}$ SPP traverses a
fast-solar-wind stream emanating from a low-latitude coronal hole, in
which $U$ and $v_{\rm A}$ are approximately steady for a period of
$\tau = 10^3 \mbox{ s}$, during which time
$U+v_{\rm A} = 1.3 \times 10^3 \mbox{ km/s}$ and
$V_{\rm sc, \perp} \simeq V_{\rm sc} = 2 \times 10^2 \mbox{ km/s}$. If
Equations~(\ref{eq:zmzp}) and (\ref{eq:vsczm}) are satisfied, then
Equations~(\ref{eqn:thHO}) and (\ref{eq:PfP3D2}) apply, and SPP
measurements during this interval sample approximately frozen~$\bm{g}$
structures along a line segment that extends from $r = 9.87 R_{\sun}$
to $r \simeq 9.87 R_{\sun} - (U+v_{\rm A})\tau = 7.99 R_{\sun}$, while
spanning a distance $V_{\rm sc} \tau \simeq 0.29 R_{\sun}$
perpendicular to the radial direction.  Over the radial range
$(7.99 R_{\sun} , 9.87 R_{\sun})$, the quantity $U+v_{\rm A}$ varies
by 6.1\% in the data-based model of \cite{Chandran:2009d}, consistent
with our assumption that $U+v_{\rm A}$ is reasonably constant
throughout this radial interval.  Given the density profile in this
model, in which $r_{\rm a} = 11.1 R_{\sun}$, Equations~(\ref{eqn:eta})
and (\ref{eqn:H}) imply that $h$ varies from a value of 0.491 at
$r=7.99 R_{\sun}$ to a value of 0.499 at $r =9.87 R_{\sun}$. To within
2\%, $h = 1/2$ throughout the interval
$(7.99 R_{\sun} , 9.87 R_{\sun})$, and thus the frequency and
wavenumber spectra of~$\bm{z}^+$ are to a high degree of accuracy
equal to a constant (1/4) times the frequency and wavenumber spectra
of~$\bm{g}$. As this example shows, the modified version of Taylor's
hypothesis that we describe will enable SPP to probe the dominant
component of solar-wind turbulence inside of SPP's perihelion.

In claiming that the energetically dominant fluctuations near the Sun
approximately satisfy Equation~(\ref{eqn:HOg3}), we have relied upon
the assumptions that: (1)
$\delta \rho / \rho_0 \ll |\delta \bm{B}| / B_0$ (which in practice
implies that $\delta \rho/\rho_0 \ll 1$ since
$|\delta \bm{B}| \lesssim B_0$); (2) $\delta \bm{B}$ and
$\delta \bm{v}$ are approximately perpendicular to $\bm{B}_0$; (3)
$z^+ \gg z^-$; (4) $V_{\rm sc,\perp} \gg z^-$; (5) $\rho_0$ varies
much more slowly with position perpendicular to~$\bm{\hat{r}}$ than
does~$\delta \bm{B}$; and (6) $\bm{B}_0$ is nearly radial.  SPP
measurements will provide tests of conditions (1) through~(5).  To check
assumption~(6) using SPP measurements would require averaging $\bm{B}$
over timescales much longer than the timescale of the energetically
dominant magnetic fluctuations, which is
likely~$\sim 1 $~h~\citep{hollweg82b}. However, such long time
averages will become progressively more difficult as SPP approaches
perihelion, since SPP will rapidly traverse the near-perihelion region
at $\sim 1 R_{\sun} \mbox{h}^{-1}$.  On the other hand, the
assumption that~$\bm{B}_0$ is nearly radial at
$r\sim 10 R_{\sun} - 30 R_{\sun}$ is likely accurate if coronal mass
ejections are excluded, because the closed-loop magnetic structures of
the Sun are primarily confined to smaller~$r$ and the Parker spiral
field begins to bend appreciably into the azimuthal direction only at
larger~$r$. We emphasize that the above conditions are sufficient but
not necessary, since, e.g., Taylor's hypothesis can apply when
$z^+ \sim z^-$ provided $z^+ \ll V_{\rm sc,\perp}$. We also note that
we have not restricted our analysis to fluctuations that are either
quasi-2D or quasi-slab-symmetric, whether with respect to the
background magnetic field~$\bm{B}_0$ or the local magnetic
field~$\bm{B}_0 + \delta \bm{B}$.

Observations at~1~AU~\citep{podesta10} and numerical simulations of
solar-wind turbulence near the Sun~\citep{Verdini:2012,Perez:2013}
suggest that if the overall fractional cross helicity is high, then
$z^+ \gg z^-$ throughout much of the inertial range.  If this is the
case near the Sun, then $\delta \bm{b} \simeq (1/2) \delta \bm{z}^+$ scale
by scale throughout much of the inertial range, and SPP magnetometer
measurements on their own (without velocity measurements) will be
sufficient to approximate the frequency spectra of $\bm{g}$
and~$\bm{z}^+$.

Our study differs from previous studies of Taylor's hypothesis for
SPP. \cite{Matthaeus:1997} accounted for $V_{\rm sc,\perp}$ and
concluded that the characteristic turbulence speed must be
$\ll V_{\rm sc,\perp}$ for Taylor's hypothesis to apply to 2D
turbulence. In contrast, we have shown that a form of Taylor's
hypothesis applies to $\bm{g}$ or $\bm{z}^+$
fluctuations even when the characteristic turbulence
speed is comparable to $V_{\rm sc,\perp}$, provided $z^- \ll z^+$ and
$z^- \ll V_{\rm sc,\perp}$.  \cite{Howes:2014a} and \cite{Klein:2014b}
found that Taylor's hypothesis holds near the perihelion of SPP
for sufficiently oblique \Alfven and kinetic \Alfven waves if the
fluctuations are treated as linear waves. In the present study, we do
not restrict the analysis to linear or highly oblique waves.

\section{Summary and Conclusion}
\label{sec:summary}

Transverse, non-compressive fluctuations likely comprise the bulk of
the fluctuation energy in the near-Sun solar wind. When
$\delta \rho \ll \rho_0$, $\bm{U}$ and $\bm{B}_0$ are nearly radial,
and $\rho_0$ varies much more slowly with position perpendicular
to~$\bm{\hat{r}}$ than does~$\delta \bm{B}$, these transverse,
non-compressive fluctuations are described by the Heinemann-Olbert
equations. We use these equations to show that outward-propagating
fluctuations can be treated as frozen within a reference frame that
will be advected past SPP at
velocity~$\bm{U} + \bm{v}_{\rm A} - \bm{V}_{\rm sc}$, provided that
$z^- \ll z^+$ and $z^- \ll V_{\rm sc,\perp}$.  Observations and
simulations suggest that these latter two inequalities and the
inequality $\delta \rho \ll \rho_0$ will be satisfied near SPP's
perihelion. As a consequence, it will likely be possible to use a
modified version of Taylor's hypothesis to characterize the spatial
structure of the outward-propagating, transverse, non-compressive
fluctuations that will be encountered by SPP near its perihelion, even
when~$U\sim v_{\rm A}$ and $z^+ \sim V_{\rm sc}$.

\acknowledgements This work was supported in part
by NSF grants AGS-1331355 and AGS-1258998, grant NNX11AJ37G from
NASA's Heliophysics Theory Program, NASA grant NNN06AA01C to the Solar
Probe Plus FIELDS Experiment, and NASA grant NNX12AB27G.



\begin{thebibliography}{}
\expandafter\ifx\csname natexlab\endcsname\relax\def\natexlab#1{#1}\fi

\bibitem[{{Barnes}(1966)}]{Barnes:1966}
{Barnes}, A. 1966, Phys.~Fluids, 9, 1483

\bibitem[{{Bavassano} {et~al.}(2000){Bavassano}, {Pietropaolo}, \&
  {Bruno}}]{Bavassano:2000}
{Bavassano}, B., {Pietropaolo}, E., \& {Bruno}, R. 2000, J.~Geophys.~Res., 105,
  15959

\bibitem[{{Bourouaine} \& {Chandran}(2013)}]{bourouaine13a}
{Bourouaine}, S., \& {Chandran}, B.~D.~G. 2013, \apj, 774, 96

\bibitem[{{Chandran}(2010)}]{Chandran:2010b}
{Chandran}, B.~D.~G. 2010, Astrophys.~J., 720, 548

\bibitem[{{Chandran} \& {Hollweg}(2009)}]{Chandran:2009d}
{Chandran}, B.~D.~G., \& {Hollweg}, J.~V. 2009, Astrophys.~J., 707, 1659

\bibitem[{{Chaston} {et~al.}(2004){Chaston}, {Bonnell}, {Carlson}, {McFadden},
  {Ergun}, {Strangeway}, \& {Lund}}]{chaston04}
{Chaston}, C.~C., {Bonnell}, J.~W., {Carlson}, C.~W., {et~al.} 2004, Journal of
  Geophysical Research (Space Physics), 109, 4205

\bibitem[{{Coleman}(1968)}]{Coleman:1968}
{Coleman}, Jr., P.~J. 1968, Astrophys.~J., 153, 371

\bibitem[{{Cranmer} \& {van Ballegooijen}(2005)}]{Cranmer:2005}
{Cranmer}, S.~R., \& {van Ballegooijen}, A.~A. 2005, Astrophys.~J.~Supp., 156,
  265

\bibitem[{{De Pontieu} {et~al.}(2007){De Pontieu}, {McIntosh}, {Carlsson},
  {Hansteen}, {Tarbell}, {Schrijver}, {Title}, {Shine}, {Tsuneta}, {Katsukawa},
  {Ichimoto}, {Suematsu}, {Shimizu}, \& {Nagata}}]{DePontieu:2007}
{De Pontieu}, B., {McIntosh}, S.~W., {Carlsson}, M., {et~al.} 2007, Science,
  318, 1574

\bibitem[{{Fredricks} \& {Coroniti}(1976)}]{Fredricks:1976}
{Fredricks}, R.~W., \& {Coroniti}, F.~V. 1976, J.~Geophys.~Res., 81, 5591

\bibitem[{{Harmon} \& {Coles}(2005)}]{Harmon:2005}
{Harmon}, J.~K., \& {Coles}, W.~A. 2005, J.~Geophys.~Res., 110, 3101

\bibitem[{{Heinemann} \& {Olbert}(1980)}]{Heinemann:1980}
{Heinemann}, M., \& {Olbert}, S. 1980, J.~Geophys.~Res., 85, 1311

\bibitem[{{Hollweg}(1978)}]{Hollweg:1978}
{Hollweg}, J.~V. 1978, Geophys.~Res.~Lett., 5, 731

\bibitem[{{Hollweg} {et~al.}(1982){Hollweg}, {Bird}, {Volland}, {Edenhofer},
  {Stelzried}, \& {Seidel}}]{hollweg82b}
{Hollweg}, J.~V., {Bird}, M.~K., {Volland}, H., {et~al.} 1982, \jgr, 87, 1

\bibitem[{{Hollweg} {et~al.}(2010){Hollweg}, {Cranmer}, \&
  {Chandran}}]{Hollweg:2010}
{Hollweg}, J.~V., {Cranmer}, S.~R., \& {Chandran}, B.~D.~G. 2010,
  Astrophys.~J., 722, 1495

\bibitem[{{Hollweg} \& {Isenberg}(2002)}]{hollweg02}
{Hollweg}, J.~V., \& {Isenberg}, P.~A. 2002, Journal of Geophysical Research
  (Space Physics), 107, 1147

\bibitem[{{Horbury} \& {Balogh}(1997)}]{horbury97}
{Horbury}, T.~S., \& {Balogh}, A. 1997, Nonlinear Processes in Geophysics, 4,
  185

\bibitem[{{Horbury} {et~al.}(2008){Horbury}, {Forman}, \&
  {Oughton}}]{horbury08}
{Horbury}, T.~S., {Forman}, M., \& {Oughton}, S. 2008, Physical Review Letters,
  101, 175005

\bibitem[{{Howes} {et~al.}(2014){Howes}, {Klein}, \& {TenBarge}}]{Howes:2014a}
{Howes}, G.~G., {Klein}, K.~G., \& {TenBarge}, J.~M. 2014, Astrophys.~J., 789,
  106

\bibitem[{{Ionson}(1978)}]{Ionson:1978}
{Ionson}, J.~A. 1978, Astrophys.~J., 226, 650

\bibitem[{{Jokipii}(1973)}]{Jokipii:1973}
{Jokipii}, J.~R. 1973, Ann.~Rev.~Astron.~Astrophys., 11, 1

\bibitem[{{Klein} {et~al.}(2014){Klein}, {Howes}, \& {TenBarge}}]{Klein:2014b}
{Klein}, K.~G., {Howes}, G.~G., \& {TenBarge}, J.~M. 2014, Astrophys.~J.~Lett.,
  790, L20

\bibitem[{{Matthaeus}(1997)}]{Matthaeus:1997}
{Matthaeus}, W.~H. 1997, Robotic Exploration Close to the Sun: Scientific Basis, 385, 67

\bibitem[{{McChesney} {et~al.}(1987){McChesney}, {Stern}, \&
  {Bellan}}]{mcchesney87}
{McChesney}, J.~M., {Stern}, R.~A., \& {Bellan}, P.~M. 1987, Physical Review
  Letters, 59, 1436

\bibitem[{{Parker}(1972)}]{Parker:1972}
{Parker}, E.~N. 1972, Astrophys.~J., 174, 499

\bibitem[{{Parker}(1988)}]{Parker:1988}
---. 1988, Astrophys.~J., 330, 474

\bibitem[{{Perez} \& {Chandran}(2013)}]{Perez:2013}
{Perez}, J.~C., \& {Chandran}, B.~D.~G. 2013, Astrophys.~J., 776, 124

\bibitem[{{Podesta} \& {Bhattacharjee}(2010)}]{podesta10}
{Podesta}, J.~J., \& {Bhattacharjee}, A. 2010, \apj, 718, 1151

\bibitem[{{Scudder}(1992)}]{Scudder:1992a}
{Scudder}, J.~D. 1992, Astrophys.~J., 398, 299

\bibitem[{{Taylor}(1938)}]{Taylor:1938}
{Taylor}, G.~I. 1938, Royal Society of London Proceedings Series A, 164, 476

\bibitem[{{Tomczyk} {et~al.}(2007){Tomczyk}, {McIntosh}, {Keil}, {Judge},
  {Schad}, {Seeley}, \& {Edmondson}}]{Tomczyk:2007}
{Tomczyk}, S., {McIntosh}, S.~W., {Keil}, S.~L., {et~al.} 2007, Science, 317,
  1192

\bibitem[{{Tu} \& {Marsch}(1995)}]{tumarsch95}
{Tu}, C., \& {Marsch}, E. 1995, Space Science Reviews, 73, 1

\bibitem[{{Velli}(1993)}]{Velli:1993}
{Velli}, M. 1993, Astron.~Astrophys., 270, 304

\bibitem[{{Velli} {et~al.}(1989){Velli}, {Grappin}, \& {Mangeney}}]{velli89}
{Velli}, M., {Grappin}, R., \& {Mangeney}, A. 1989, Physical Review Letters,
  63, 1807

\bibitem[{{Verdini} {et~al.}(2012){Verdini}, {Grappin}, {Pinto}, \&
  {Velli}}]{Verdini:2012}
{Verdini}, A., {Grappin}, R., {Pinto}, R., \& {Velli}, M. 2012,
  Astrophys.~J.~Lett., 750, L33

\bibitem[{{Verdini} \& {Velli}(2007)}]{Verdini:2007}
{Verdini}, A., \& {Velli}, M. 2007, Astrophys.~J., 662, 669

\end{thebibliography}


\end{document}